\documentstyle[preprint,aps,epsfig]{revtex}
\newcommand{\eq}{\begin{eqnarray}}
\newcommand{\en}{\end{eqnarray}}
\tightenlines
\begin{document}

\title{Electromagnetic nucleon-delta transition in the\\
perturbative chiral quark model}

\author{K.\ Pumsa-ard, V.\ E.\ Lyubovitskij,
Th.\ Gutsche, Amand Faessler \\ and S.\ Cheedket\footnote{Present address: School of Physics,
Institute of Science, Suranaree University of Technology, Nakhon Ratchasima 30000, Thailand}
\vspace*{0.4\baselineskip}}
\address{
Institut f\"ur Theoretische Physik, Universit\"at T\"ubingen,\\
Auf der Morgenstelle 14, D-72076 T\"ubingen, Germany
\vspace*{0.3\baselineskip}\\}

\maketitle

\vskip.5cm

\begin{abstract}
We apply the perturbative chiral quark model to the $ \gamma N \to \Delta $ transition.
The four momentum dependence of the respective transverse helicity
amplitudes $A_{1/2}$ and $A_{3/2}$ is determined at one loop in the
pseudoscalar Goldstone boson
fluctuations. Inclusion of excited states in the quark propagator is shown
to result in a reasonable description of the experimental values for the
helicity amplitudes at the real photon point.
\end{abstract}

\vskip1cm


\vskip.5cm


\newpage

\section{Introduction}

Electromagnetic transitions of the nucleon to baryon excitations
give important insight into the degrees of freedom which are relevant
in hadron physics and hence for the structure of the nucleon.
From this point of view the study of the particular transition $\gamma N \to
\Delta(1232)$ is sensitive to the spatial and spin structure
of the involved baryons.

A comparison between theory and experiment is conveniently performed
on the level of electromagnetic transition matrix elements, which are
expressed in terms of photon helicity amplitudes \cite{Copley,PDG}.
For transverse photons the electromagnetic $N-\Delta$ transition is described by the
helicity amplitudes $A_{\lambda}$, where $\lambda = 3/2 (1/2)$ refers
to the case where the photon spin is parallel (antiparallel) to the spin
of the nucleon target. In turn, the decay rate for $\Delta^+ \to p \gamma$
can be expressed in terms of these transverse helicity amplitudes \cite{PDG,Jenkins}
\eq\label{dwidth}
\Gamma(\Delta^+ \to p \gamma) = \frac{(P^\ast)^2}{2\pi} \,
\biggl(\frac{M_p}{M_\Delta}\biggr) \, \{ |A_{1/2}|^2 + |A_{3/2}|^2 \}\,,
\en
where $M_p$ and $M_\Delta$ are the baryon masses and
\eq
P^\ast = \frac{M_\Delta^2 - M_p^2}{2M_\Delta} = 258.4 \,\,
{\rm MeV}
\en
is the 3-momentum of either of the final state particles
(proton or real photon) in the $\Delta$-rest frame.
Experimental values for the helicity amplitudes $A_{1/2}$ and $A_{3/2}$ at the
real-photon point and for
the branching ratio ${\rm BR}(\Delta^+ \to p \gamma) =
\Gamma (\Delta^+ \to p \gamma)/\Gamma_{total}(\Delta^+)$ are reported as \cite{PDG}
\eq
&&A_{1/2}=-(135 \pm 6) \times 10^{-3} {\rm GeV}^{-1/2}, \hspace*{1cm}
A_{3/2}=-(255 \pm 8) \times 10^{-3} {\rm GeV}^{-1/2}, \\
&&\frac{A_{3/2}}{A_{1/2}} = 1.89 \pm 0.10, \hspace*{3.8cm}
{\rm BR}(\Delta^+ \to p \gamma) = (0.52 - 0.60) \%.
\nonumber
\en
Alternatively, the transverse helicity amplitudes can be expressed in
terms of electromagnetic production multipoles, that is the magnetic
dipole $M1$ and the electric quadrupole $E2$ moments. The two sets of
amplitudes are related by
\eq
A_{1/2} = - \frac{1}{2} (M1 + 3 E2), \hspace*{2cm}
A_{3/2} = - \frac{\sqrt{3}}{2} (M1 - E2)\, .
\en
Thereby, a non-vanishing $E2/M1$ ratio indicates the possibility of an intrinsic
deformation of the nucleon \cite{IsgurKarl,Gershtein,Drechsel} or
a contribution from meson exchange currents \cite{Buchmann_EM,Buchmann_EC} with the latest
experimental value of \cite{PDG}
\eq\label{emratio}
\frac{E2}{M1} = - 0.025 \pm 0.005 .
\en

Various model calculations have been applied to the transverse helicity
amplitudes of the electromagnetic $N \to \Delta$ transition.
In the original constituent quark model~\cite{Copley,Isgur,Giannini} 
the results for the helicity amplitudes at the real-photon point are somewhat smaller 
than the experimental data. Further extensions of the constituent quark model by 
considering two-body currents based on gluon and exchange were performed 
in \cite{Buchmann_EM,Buchmann_EC}. However, in these calculations lower values for 
the $M1$ contribution as compared to the experimental ones are still obtained even 
when also including relativistic effects. In the cloudy
bag model \cite{KE,Bermuth,Lu} this situation was improved where pion cloud
corrections are shown to contribute up to two thirds of the total amplitude.
Similar models, e.g., the relativistic potential quark model \cite{Dong}, confirmed 
the importance of meson cloud corrections.
In these model calculations one is also able to generate a non-vanishing
value for the $E2/M1$ ratio, which is extremely sensitive to
non-valence quark degrees of freedom, referred to as the pion cloud of the
nucleon or as exchange currents. With the exception of Ref.~\cite{Buchmann_EM},
predictions for this ratio are usually considerably smaller than the 
experimental value of Eq. (\ref{emratio}). 
A significant contribution from the meson cloud is also emphasized
in Ref.~\cite{Sato} where the dynamical pion cloud effects
originate from the non-resonant $\pi$ production mechanism.
These effects are crucial in getting agreement with the data.
The improvement of the theoretical calculations arising from the
meson cloud effects is not only restricted to the $\gamma N \to \Delta$
transition but also applies to other transitions as well, e.g., weak pion
production reactions \cite{Sato}, vector, axial-vector and strong $NN$
and $N\Delta$ couplings \cite{Hemmert} and light meson
photoproduction \cite{Kamalov}.
 
As a further development of chiral quark models with a perturbative treatment
of the Goldstone boson cloud \cite{Theberge}-\cite{Chin}
we recently developed the perturbative chiral quark
model (PCQM) for the description of low-energy properties of baryons
\cite{Gutsche}-\cite{PCQM4}.
The PCQM is based on the nonlinear $\sigma$-model quark
Lagrangian and includes a phenomenological confinement potential.
Baryons are considered as bound states of valence quarks surrounded
by a cloud of pseudoscalar mesons as imposed by chiral symmetry
requirements.
The model was successfully applied to the electromagnetic properties of the
nucleon \cite{PCQM1}, $\sigma$-term physics \cite{PCQM2}, the $\pi N$
scattering including radiative corrections \cite{PCQM3} and the strange
nucleon form factors \cite{PCQM4}.

In the current work we consider, as a further extension, the determination
of the momentum dependence of the helicity amplitudes $A_{1/2}$ and $A_{3/2}$
of the $N-\Delta$ transition at one-loop or equivalently to the order of accuracy
$o(1/F^2,\hat m,m_s)$. Here, F is the pion decay constant in the
chiral limit and $\hat m, m_s$ are the respective current masses of
up/down and strange quarks. In the context of the current approach we furthermore
investigate the role of excited quark states in meson loop diagrams.
Whereas in our previous considerations the expansion of the quark propagator was
restricted to include only ground state contributions, that is N and $\Delta$ intermediate
states in loop diagrams, we now also consider excited quark states.
The main conclusion drawn from this calculation will be that inclusion of excited
quark states are relevant at the $20\%$ level in order to obtain a
reasonable description of the transverse helicity amplitudes at the real-photon point.
For the $E2/M1$ ratio we trivially obtain a vanishing value, since at the order
of accuracy we are working in a non-vanishing result cannot be expected due to
large-$N_c$ QCD arguments \cite{Jenkins}.

In the present article we proceed as follows. In the following section
we review the basic notions of the perturbative chiral quark model and the modifications
of the perturbative technique when including the full quark propagator.
In Sec. III we present the calculation of the transverse
$N- \Delta $ helicity amplitudes and give a comparison with current experimental data.
Finally, Sec.IV contains a summary of our major conclusions.

\section{The Perturbative Chiral Quark Model (PCQM)}

The starting point of the perturbative chiral quark
model~\cite{Gutsche}-\cite{PCQM4} is an effective chiral Lagrangian
describing the valence quarks of baryons as relativistic fermions
moving in an external field (static potential) $V_{\rm eff}(r)=S(r)+\gamma^0 V(r)$
with $r=|\vec x|$ \cite{Gutsche,PCQM1}, which in the SU(3)-flavor version
are supplemented by a cloud of Goldstone bosons $(\pi, K, \eta)$.
Treating Goldstone fields as small fluctuations around the
three-quark (3q) core we have the linearized effective Lagrangian \cite{PCQM1}:
\eq\label{linearized_L}
{\cal L}_{\rm eff}(x) &=&
\bar{\psi}(x) [i \not\!\partial - S(r) - \gamma^0 V(r)]\psi(x) +
\frac{1}{2} [\partial_\mu \hat \Phi(x)]^2  \nonumber \\
 &-& \bar{\psi}(x) S(r) i \gamma^5 \frac{\hat \Phi (x)}{F}
 \psi(x)+{\cal L}_{\chi SB}(x).
\en
The additional term ${\cal L}_{\chi SB}$ contains the mass contributions both for quarks
and mesons, which explicitly break chiral symmetry:
\begin{equation}
{\cal L}_{\chi SB}(x) = -\bar\psi(x) {\cal M} \psi(x)
- \frac{B}{2} Tr [\hat \Phi^2(x)  {\cal M} ].
\end{equation}
Here, $\hat \Phi$ is the octet matrix of pseudoscalar mesons
\eq
\frac{\hat \Phi}{\sqrt{2}} = \left( \begin{array}{ccc}
\pi^0/\sqrt{2}+\eta/\sqrt{6} & \pi^+ & K^+ \\
\pi^- & -\pi^0/\sqrt{2} + \eta/\sqrt{6} & K^0 \\
K^- & \bar K^0  & -2\eta/\sqrt{6}
\end{array} \right)\, ,
\en
$F=88$ MeV is the pion decay constant in the chiral
limit \cite{Gasser_Sainio}, ${\cal M}={\rm diag}\{\hat m,\hat m,m_s\}$
is the mass matrix of current quarks (we restrict to the isospin
symmetry limit $m_u=m_d=\hat m$) and $B=-<0|\bar u u|0>/F^2$ is the quark
condensate constant. We rely on the standard picture of chiral symmetry
breaking~\cite{Gasser_Leutwyler} and for the masses of pseudoscalar
mesons we use the leading term in their chiral expansion (i.e. linear
in the current quark mass):
\eq\label{M_Masses}
M_{\pi}^2=2 \hat m B, \hspace*{.5cm} M_{K}^2=(\hat m + m_s) B,
\hspace*{.5cm} M_{\eta}^2= \frac{2}{3} (\hat m + 2m_s) B.
\en
The following set of
parameters~\cite{Gasser_Leutwyler} is chosen in our evaluation
\begin{equation}
\hat m = 7 \;{\rm MeV},\; \frac{m_s}{\hat m}=25,\;
B = \frac{M^2_{\pi^+}}{2 \hat m}=1.4 \;{\rm GeV}.
\end{equation}
Meson masses obtained in Eq.~(\ref{M_Masses}) satisfy the
Gell-Mann-Oakes-Renner and the Gell-Mann-Okubo relation. In addition,
the linearized effective Lagrangian in Eq.~(\ref{linearized_L})
fulfils the PCAC requirement.

We expand the quark field $\psi$ in the basis of potential
eigenstates as
\eq\label{total_psi}
\psi(x) = \sum\limits_\alpha b_\alpha u_\alpha(\vec{x})
\exp(-i{\cal E}_\alpha t) + \sum\limits_\beta
d_\beta^\dagger v_\beta(\vec{x}) \exp(i{\cal E}_\beta t)\, ,
\nonumber
\en
where the sets of quark $\{ u_\alpha \}$ and antiquark $\{ v_\beta \}$
wave functions in orbits $\alpha$ and $\beta$ are solutions of the
Dirac equation with the static potential $V_{\rm eff}(r)$.
The expansion coefficients $b_\alpha$ and $d_\beta^\dagger$ are the
corresponding single quark annihilation and antiquark creation
operators.

We formulate perturbation theory in the expansion parameter
$1/F$ ($F \sim \sqrt{N_c})$ and treat finite current quark
masses perturbatively \cite{PCQM1}. All calculations are performed
at one loop or at order of accuracy $o(1/F^2, \hat{m}, m_s)$.
In the calculation of matrix elements we project quark diagrams on
the respective baryon states. The baryon states are conventionally
set up by the product of the ${\rm SU(6)}$ spin-flavor and
${\rm SU(3)_c}$ color wave functions, where the nonrelativistic single
quark spin wave function is replaced by the relativistic solution
$u_\alpha(\vec{x})$ of the Dirac equation
\begin{equation}\label{Dirac_eq}
\left[ -i\gamma^0\vec{\gamma}\cdot\vec{\nabla} + \gamma^0 S(r) + V(r)
- {\cal E}_\alpha \right] u_\alpha(\vec{x})=0,
\end{equation}
where ${\cal E}_\alpha$ is the single-quark energy.

For the description of baryon properties we use the effective potential
$V_{\rm eff}(r)$ with a quadratic radial dependence \cite{PCQM1,PCQM2}:
\eq\label{V_eff}
S(r) = M_1 + c_1 r^2, \hspace*{1cm} V(r) = M_2+ c_2 r^2
\en
with the particular choice
\eq
M_1 = \frac{1 \, - \, 3\rho^2}{2 \, \rho R} , \hspace*{1cm}
M_2 = {\cal E}_0 - \frac{1 \, + \, 3\rho^2}{2 \, \rho R} , \hspace*{1cm}
c_1 \equiv c_2 =  \frac{\rho}{2R^3} .
\en
Here, ${\cal E}_0$ is the single-quark ground-state energy;
$R$ are $\rho$ are parameters related to the ground-state quark wave
function $u_0$:
\eq\label{Gaussian_Ansatz}
u_0(\vec{x}) \, = \, N \, \exp\biggl[-\frac{\vec{x}^{\, 2}}{2R^2}\biggr]
\, \left(
\begin{array}{c}
1\\
i \rho \, \vec{\sigma}\vec{x}/R\\
\end{array}
\right)
\, \chi_s \, \chi_f \, \chi_c,
\en
where $N=[\pi^{3/2} R^3 (1+3\rho^2/2)]^{-1/2}$ is a normalization
constant; $\chi_s$, $\chi_f$, $\chi_c$ are the spin, flavor and color
quark wave function, respectively. Note, that the constant part of the
scalar potential $M_1$ can be interpreted as the constituent mass of
the quark, which is simply the displacement of the current quark mass
due to the potential $S(r)$. The parameter $\rho$ is related to the
axial charge $g_A$ of the nucleon calculated in zeroth-order
(or 3q-core) approximation:
\eq\label{ga_rho_match}
g_A=\frac{5}{3}\biggl(1 - \frac{2\rho^2}{1+\frac{3}{2}\rho^2}\biggr)\,.
\en
Therefore, $\rho$ can be replaced by $g_A$ using the matching condition
(\ref{ga_rho_match}). The parameter $R$ is related to the charge radius
of the proton in the zeroth-order approximation as
\eq\label{rad_LO}
<r^2_E>^P_{LO} = \int d^3 x \, u^\dagger_0 (\vec{x}) \,
\vec{x}^{\, 2} \, u_0(\vec{x}) \, = \, \frac{3R^2}{2} \,
\frac{1 \, + \, \frac{5}{2} \, \rho^2}{1 \, + \, \frac{3}{2} \, \rho^2}.
\en
In our calculations we use the value $g_A$=1.25 \cite{PCQM1}. Therefore,
we have only one free parameter, that is $R$. In the numerical
studies \cite{PCQM1} R is varied in the region from 0.55 fm to 0.65 fm,
which corresponds to a change of $<r^2_E>^P_{LO}$
from 0.5 to 0.7 fm$^2$.

The expectation value of an operator $\hat A$
is set up as:
\begin{equation}\label{perturb_A}
<\hat A> = {}^B\!\!< \phi_0 |\sum^{\infty}_{n=1} \frac{i^n}{n!}
\int d^4 x_1 \ldots \int d^4 x_n T[{\cal L}_I (x_1) \ldots
{\cal L}_I (x_n) \hat A]|\phi_0>^B_c,
\end{equation}
where the state vector $|\phi_0>$ corresponds to the unperturbed
three-quark state ($3q$-core).
Superscript $``B"$ in (\ref{perturb_A}) indicates that the
matrix elements have to be projected onto the respective
baryon states, whereas subscript $``c"$ refers to contributions from
connected graph only. ${\cal L}_I (x)$ of Eq. (\ref{perturb_A})
refers to the linearized quark-meson interaction Lagrangian:
\begin{equation}
{\cal L}_I (x) = -\bar \psi (x) i \gamma^5 \frac{\hat \Phi (x)}{F}
S(r) \psi(x).
\end{equation}
For the evaluation of Eq.(\ref{perturb_A}) we apply Wick's
theorem with the appropriate propagators for quarks and mesons.

For the quark field we use a Feynman
propagator for a fermion in a binding potential with
\eq\label{quark_propagator}
i G_\psi(x,y) &=& <\phi_0|T\{\psi(x)\bar \psi(y)\}|\phi_0>\nonumber \\
&=& \theta(x_0-y_0) \sum\limits_{\alpha} u_\alpha(\vec{x})
\bar u_\alpha(\vec{y}) e^{-i{\cal E}_\alpha (x_0-y_0)}
- \theta(y_0-x_0) \sum\limits_{\beta} v_\beta(\vec{x})
\bar v_\beta(\vec{y}) e^{i{\cal E}_\beta (x_0-y_0)} .
\en
In previous applications \cite{PCQM1}-\cite{PCQM4} we restricted
the expansion of the quark propagator to its ground state with:
\eq\label{quark_propagator_ground}
iG_\psi(x,y) \to iG_0(x,y) \doteq u_0(\vec{x}) \, \bar u_0(\vec{y}) \,
e^{-i{\cal E}_0 (x_0-y_0)} \, \theta(x_0-y_0).
\en
Such a truncation can be considered as an additional regularization of
the quark propagator, where in the case of SU(2)-flavor intermediate baryon states
in loop-diagrams are restricted to $N$ and $\Delta$.
In the current approach we also include, for the first time,
excited quark states in the propagator of Eq.~(\ref{quark_propagator}) and
analyse their influence on the matrix elements for the $N$-$\Delta$ transitions
considered.
We include the following set of excited quark states: the first
$p$-states ($1p_{1/2}$ and $1p_{3/2}$ in the non-relativistic notation) and the
second excited states ($1d_{3/2}, 1d_{5/2}$ and $2s_{1/2}$).
For the given form of the effective potential (\ref{V_eff}) the Dirac
equation can be solved analytically.
The corresponding expressions for the wave functions of the excited quark states
are given in the Appendix.
For the meson fields we adopt the free Feynman propagator with
\eq
i\Delta_{ij}(x-y)=<0|T\{\Phi_i(x)\Phi_j(y)\}|0>=\delta_{ij}
\int\frac{d^4k}{(2\pi)^4i}\frac{\exp[-ik(x-y)]}{M_\Phi^2-k^2-i\epsilon}.
\en

Introduction of the electromagnetic field $A_\mu$ to the effective
Lagrangian (\ref{linearized_L}) is accomplished by minimal substitution:
\eq
\partial_\mu\psi \to D_\mu\psi = \partial_\mu\psi + i e Q A_\mu \psi,
\hspace*{.5cm}
\partial_\mu\Phi_i \to D_\mu\Phi_i = \partial_\mu\Phi_i +
e \biggl[f_{3ij} + \frac{f_{8ij}}{\sqrt{3}} \biggr] A_\mu \Phi_j
\en
where $Q = {\rm diag}\{2/3, -1/3, -1/3\}$ is the quark charge matrix
and $f_{ijk}$ are the totally antisymmetric structure constants of
$SU(3)$.
For the photon field $A_\mu$ we also include the usual kinetic term
\eq\label{L_ph}
{\cal L}_{ph} = - \frac{1}{4} F_{\mu\nu}(x)F^{\mu\nu}(x) \hspace*{.5cm}
\mbox{with} \hspace*{.5cm} F_{\mu\nu}(x) = \partial_\nu A_\mu(x) -
\partial_\mu A_\nu(x) .
\en

In the evaluation of Eq. (\ref{perturb_A})
we redefine the perturbation series in terms of
renormalized quantities, where a set of counterterms $\delta{\cal L}$
has to be introduced in the Lagrangian (\ref{linearized_L}).
The counterterms play a dual role:
i) they maintain the proper definition of physical parameters, such as
nucleon mass and, in particular, the nucleon charge and ii) they
effectively reduce the number of Feynman diagrams to be evaluated.
For a detailed derivation and discussion of this renormalization
technique see the original reference \cite{PCQM1}.
Here we just indicate the relevant results following the technique
of Ref. \cite{PCQM1}, where now intermediate excited quark states are included
in the loop diagrams.
In the following we attach the index $``0"$ to quantities
when we truncate the quark propagator to the ground state
contribution, index $``F"$ refers to the case where
excited states are also included.

First we introduce the renormalized quark field $\psi^r(x)$, which
is a solution to the Dirac equation (\ref{Dirac_eq})
with the full, renormalized non-strange quark mass \cite{PCQM1}
\begin{equation}\label{mrenorm}
\hat m^r_F \, = \, \hat m \, - \, \sum\limits_{\alpha} \,
\frac{1}{3\gamma} \, \biggl(\frac{1}{\pi F}\biggr)^2
\int\limits_0^\infty dp \,p^2 \, F_{\alpha}(p^2) \,
F^{\dagger}_{\alpha}(p^2) \biggl\{ \frac{9}{4}
{\cal C}^{\pi}_{\alpha}(p^2) + \frac{3}{2} {\cal C}^{K}_{\alpha}(p^2) +
\frac{1}{4} {\cal C}^{\eta}_{\alpha}(p^2)\biggr\},
\end{equation}
including self-energy corrections of the meson cloud.
In Eq. (\ref{mrenorm}) we introduce the relativistic reduction factor
$\gamma=(1-\frac{3}{2}\rho^2)/(1+\frac{3}{2}\rho^2)$ and
\begin{equation}
{\cal C}^{\phi}_{\alpha}(p^2)=\frac{1}{w_{\phi}(p^2) \,
(w_{\phi}(p^2)+\Delta{\cal E}_{\alpha})}
\end{equation}
with meson energy $w_\Phi(p^2)=\sqrt{M_\Phi^2+p^2}$ and momentum $p=|\vec{p}\,|$.
$\Delta {\cal E}_{\alpha} = {\cal E}_{\alpha} - {\cal E}_0$ is
the excess of the energy of the quark in state $\alpha$ with respect to the
ground state. At the quark-meson vertex we obtain the form factor for
the transition from a quark in the ground state to an excited one
($\alpha=(nljm)$) with
\eq\label{F-ex}
F_{\alpha}(p^2) &=& N N_{\alpha} \frac{\partial}{\partial p}
\int\limits^\infty_0 dr \,r\,
S(r) (g_0(r) f_{\alpha}(r) + g_{\alpha}(r) f_0(r)) \nonumber \\
&\times&\int\limits_{\Omega} \, d\cos\theta \, d\phi \,
e^{i p r cos \theta} \, C_\alpha \, Y_{l\,0}(\theta,\phi)
\en
with $C_{\alpha} \doteq (l\,0\,\frac{1}{2}\,\frac{1}{2}|j\,\frac{1}{2})$.
Explicit forms of the radial wave functions ($g_{\alpha}$
and $f_{\alpha}$), normalizations ($N_{\alpha}$) and energy difference
($\Delta{\cal E}_{\alpha}$) are given in the Appendix.
When the quark propagator is restricted to the ground state only, the
renormalized quark mass reduces to \cite{PCQM1}
\begin{equation}
\hat m^r_0=\hat m-\frac{1}{3 \gamma}\frac{9}{400}\biggl( \frac{g_A}{\pi F}
\biggr)^2\int\limits_0^\infty dp\,p^4\,F^2_{\pi NN}(p^2)\biggl\{
\frac{9}{w^2_{\pi}(p^2)} + \frac{6}{w^2_{K}(p^2)}
+ \frac{1}{w^2_{\eta}(p^2)}\biggr\},
\end{equation}
where $F_{\pi NN}(p^2)$ is the $\pi NN$
form factor normalized to unity at zero recoil $(p^2=0)$:
\eq\label{F_piNN}
F_{\pi NN}(p^2) = \exp\biggl(-\frac{p^2R^2}{4}\biggr)
\biggl\{ 1 \, + \, \frac{p^2R^2}{8}
\biggl(1 \, - \, \frac{5}{3g_A}\biggr)\biggr\} .
\en
For the effective potential considered
the relationship between renormalized and bare potential eigenstates
can be deduced analytically.

When the original Lagrangian (\ref{linearized_L}) is rewritten in terms
of renormalized quark fields and masses, a set of
counterterms, denoted by $\delta {\cal L}^{str}$, has to be included:
\eq\label{counter_terms}
\delta {\cal L}^{str}&=& \delta {\cal L}^{str}_1 +
\delta {\cal L}^{str}_2 + \delta {\cal L}^{str}_3,
\label{L_ct_all}\nonumber\\
\mbox{with}& &\nonumber\\
\delta {\cal L}^{str}_1&=& \bar\psi^r(x) \, (Z - 1) \,
[i\not\! \partial - {\cal M}^r - S(r) - \gamma^0V(r)]\psi^r(x),
\label{L_ct_str1}\nonumber\\
& &\nonumber\\
\delta {\cal L}^{str}_2&=& - \, \bar\psi^r(x) \, \delta {\cal M} \,
\psi^r(x), \label{L_ct_str2}\nonumber\\[3mm]
\delta {\cal L}^{str}_3&=& - c_\pi \sum\limits_{i=1}^{3}
[\bar\psi^r(x) i\gamma^5 \lambda_i\psi^r(x)]^2  -
c_K \sum\limits_{i=4}^{7} [\bar\psi^r(x) i\gamma^5
\lambda_i\psi^r(x)]^2 - c_\eta [\bar\psi^r(x)
i\gamma^5 \lambda_8\psi^r(x)]^2. \label{L_ct_str4}
\en
In the counterterms ${\cal M}^r$ is the mass matrix of renormalized quark masses
and ${\cal M}^r = {\cal M} -  \delta {\cal M}$.
The terms in $\delta {\cal L}^{str}_3$ are introduced for the
purpose of nucleon mass renormalization due to meson exchange between
different quarks and contain the factors
\eq
c_\Phi \, =  \, - \frac{9}{200} \, \frac{(2\pi R^2)^{3/2}}{(1-\gamma^2)} \,
\Pi (M_\Phi^2) \, \,
\mbox{with} \, \,
\Pi (M_\Phi^2) \, =  \, - \, \biggl(\frac{g_A}{\pi F}\biggr)^2 \,\,
\int\limits_0^\infty \frac{dp \, p^4}{w_\Phi^2(p^2)} \,\, F_{\pi NN}^2(p^2) .
\en
The renormalization constants $\hat Z$ for $u,d$-quarks and $Z_s$
for the $s$-quark are given in the diagonal matrix
$Z = {\rm diag} \{\hat Z, \hat Z, Z_s \}$.
The values of $\hat Z$ and $Z_s$ are determined by the
charge conservation condition at one loop \cite{PCQM1}. Due to the consistency of
our approach the values of $\hat Z$ and $Z_s$ guarantee charge conservation
both on the quark and the baryon level.
In the context of the present calculation
we only have to consider the renormalization constant $\hat Z$
for the non-strange quarks. When excited quark states are included we obtain
the following analytical expression for the renormalization constant
\begin{equation}\label{Z-F}
\hat Z^F = 1-\sum_{\alpha}\biggl( \frac{1}{\pi F} \biggr)^2\int
\limits_0^\infty dp \,p^2 \,\, F_{\alpha}(p^2) F^{\dagger}_{\alpha}(p^2)
\biggl\{\frac{3}{4}{\cal W}^{\pi}_{\alpha}(p^2)+\frac{1}{2}{\cal
W}^{K}_{\alpha}(p^2)+\frac{1}{12}{\cal W}^{\eta}_{\alpha}(p^2) \biggr\}
\end{equation}
with
\begin{equation}
{\cal W}^{\phi}_{\alpha}(p^2)=
\frac{1}{w_{\phi}(p^2)(w_{\phi}(p^2)+\Delta{\cal E}_{\alpha})^2}.
\end{equation}
When we restrict intermediate quark states to the ground state,
the expression of Eq.(\ref{Z-F}) reduces to
\eq\label{Z-eq}
\hat Z^0 = 1 - \, \frac{27}{400} \biggl(\frac{g_A}{\pi F}\biggr)^2
\int\limits_0^\infty dp \, p^4 \,\, F_{\pi NN}^2(p^2)
\biggl\{ \frac{1}{w_\pi^3(p^2)} + \frac{2}{3w_K^3(p^2)}
+ \frac{1}{9w_\eta^3(p^2)}  \biggr\} .
\en
which has already been derived in \cite{PCQM1}.
In the two-flavor picture, that is when we restrict to
the pion cloud contribution only, we obtain a value of
$\hat Z^0 = 0.9 \pm 0.02$ for our set of parameters.
The contribution of kaon and $\eta$-meson loops to the constant
$\hat Z^0$ is strongly suppressed due to the energy denominators
in Eq. (\ref{Z-eq}). In the three-flavor picture we get
$\hat Z^0=0.88 \pm 0.03$, which deviates only slightly from the
two-flavor result. With the value for $\hat Z^0$ being close to unity,
for our set of parameters the perturbative treatment of the meson
cloud is also justified. Inclusion of the excited states changes the
value of the renormalization constant to $\hat Z^F=0.81 \pm 0.04$.

For the $N \to \Delta $ transition
we use the renormalized electromagnetic current operator
$j^\mu_r$ as derived by Noether's theorem \cite{PCQM1}:
\eq\label{em_current}
j^\mu_r = j^\mu_{\psi^r} \, + \, j^\mu_\Phi \, + \delta j^\mu_{\psi^r} .
\en
It contains the quark component $j^\mu_{\psi^r}$, the charged meson
component $j^\mu_\Phi$ and the contribution of the counterterm
$\delta j^\mu_{\psi^r}$:
\eq
j^\mu_{\psi^r} &=& \bar\psi^r \gamma^\mu Q \psi^r
\equiv \frac{1}{3} \,\,
[ 2 \, \bar u^r \gamma^\mu u^r - \bar d^r \gamma^\mu d^r
- \bar s^r \gamma^\mu s^r] ,\label{em_psi_currents}\\
j^\mu_\Phi &=& \biggl[f_{3ij} + \frac{f_{8ij}}{\sqrt{3}}\biggr] \Phi_i
\partial^\mu \Phi_j \equiv \,\, [\pi^- i\partial^\mu \pi^+ -
\pi^+ i\partial^\mu \pi^- \, + \, K^- i\partial^\mu K^+ -
K^+ i\partial^\mu K^-],\label{em_psi_Phi_currents}\nonumber\\
\mbox{and}& &\nonumber\\
\delta j^\mu_{\psi^r} &=&  \bar\psi^r \, (Z - 1) \, \gamma^\mu Q \psi^r
\equiv \frac{1}{3} \,
[ 2 \, (\hat{Z} - 1) \, \bar u^r \gamma^\mu u^r - (\hat{Z} - 1) \,
\bar d^r \gamma^\mu d^r  - (Z_s - 1) \bar s^r \gamma^\mu s^r] .
\label{em_ct_currents} \nonumber
\en
Again, further technical details of the presented formalism are contained in the
original reference \cite{PCQM1}, but here it is extended to the case where
excited states are included in the quark propagator.

\section{$\gamma N \to \Delta$ helicity amplitudes in the PCQM}

Now we consider the determination of the helicity amplitudes
for the transition $\gamma N \to \Delta$ in the PCQM.
The transverse helicity amplitudes are defined as
\begin{equation}
A_M = - \frac{e}{\sqrt{2 \omega_\gamma}} <\Delta,s^{'}_{z} =
M | \vec{j} \cdot \vec{\epsilon} \, | N,s_z = M-1>
\end{equation}
where $M = \frac{1}{2}, \frac{3}{2}$ and
$\omega_\gamma \equiv P^\ast = (M^2_{\Delta}-M^2_N)/2M_{\Delta}$
is the energy of the photon in the rest frame of the $\Delta$ with
the polarization vector $\vec \epsilon$; $M_{N}$ and $M_{\Delta}$ are
the physical masses of the nucleon and the $\Delta(1232)$-resonance,
respectively.

In the PCQM the helicity amplitudes $A_{1/2}(q^2)$ and $A_{3/2}(q^2)$,
where $q$ is the photon four-momentum with $Q^2 = - q^2$, are then
identified with the perturbative expressions:
\eq
A_{1/2}(Q^2) &=& - \frac{e}{\sqrt{2 \omega_\gamma}}<\Delta^+ \, , 1/2| \, - \frac{1}{2} \,
\int \, \delta(t) \, d^4x \, d^4x_1 \, d^4x_2 \, e^{-iqx} \,
\nonumber\\
&\times& T[{\cal L}_{str}^r(x_1) \, {\cal L}_{str}^r(x_2) \,
\vec{j}_r(x) \cdot \vec\epsilon \, ] \, |p \, , -1/2>_c , \label{A12}\\
A_{3/2}(Q^2) &=& - \frac{e}{\sqrt{2 \omega_\gamma}}
<\Delta^+ \, , 3/2| \, - \frac{1}{2} \,
\int \, \delta(t) \, d^4x \, d^4x_1 \, d^4x_2 \, e^{-iqx} \,
\nonumber\\
&\times& T[{\cal L}_{str}^r(x_1) \, {\cal L}_{str}^r(x_2) \,
\vec{j}_r(x) \cdot \vec\epsilon \, ] \, |p \, , 1/2>_c \, ,
\label{A32}
\en
where $|B\, , M>  \doteq |\phi_0>^{B_M}$ refers to the unperturbed
three-quark state of baryon B with spin projection M.
Here, $\vec{j}_r$ is the spatial part of the renormalized
electromagnetic current operator already introduced in Eq. (\ref{em_current}).
${\cal L}_{str}^r = {\cal L}^{str}_I + \delta {\cal L}^{str}$
is the renormalized strong interaction Lagrangian \cite{PCQM1} containing the
quark-meson interaction term
\eq\label{L_str_I}
{\cal L}^{str}_I = - \bar\psi^r(x) i\gamma^5 \frac{\hat\Phi(x)}{F}
S(r) \psi^r(x)
\en
and the set of counterterms $\delta {\cal L}^{str}$ of Eq. (\ref{counter_terms}).

The helicity amplitudes $A_{1/2}$ and $A_{3/2}$ are evaluated at one-loop
or to the order of accuracy $o(1/F^2,\hat m,m_s)$.
At this level, which is also equivalent to $O(1/N_c)$ as will be discussed
later, we obtain the naive
relation $A_{3/2} = \sqrt{3} \cdot A_{1/2}$, which can also be
checked explicitly. Therefore, in the following we only present explicit analytical results
for the helicity amplitude $A_{1/2}$.

We start with the simplest case, where the quark propagator is restricted to the
ground state contribution.
The total helicity amplitude is a sum of terms arising from different diagrams:
the three-quark diagram (Fig.1a), the counterterm (Fig.1b),
the meson-cloud diagram (Fig.1c), the vertex-correction diagram (Fig.1d) and the
meson-in-flight diagram
(Fig.1e).

(a) The helicity amplitude due to the three-quark diagram [Fig.1(a)] is given by a
leading order (LO) and a next-to-leading (NLO) term, where the latter one
arises after renormalization \cite{PCQM1}:
\eq
A_{1/2}(Q^2)\bigg|_{3q} &=& A_{1/2}(Q^2)\bigg|^{LO}_{3q} +
A_{1/2}(Q^2)\bigg|^{NLO}_{3q} \\
A_{1/2}(Q^2)\bigg|^{LO}_{3q} &=& -\frac{2}{3}
\frac{e \; P^\ast(Q^2)}{\sqrt{2\omega_\gamma}} \rho R \frac{{\rm exp}
\Big( -\frac{Q^2 R^2}{4} \Big) }{1 + \frac{3\rho^2}{2}}
\en
\begin{equation}\label{A12NLO_0}
A_{1/2}(Q^2)\bigg|^{NLO}_{3q} = A_{1/2}(Q^2)\bigg|^{LO}_{3q}
\hat m^r_0\Bigg(\frac{\rho R}{1 + \frac{3 \rho^2}{2}} \Bigg)
\Bigg(\frac{Q^2 R^2}{4} - \frac{2 -
\frac{3 \rho^2}{2}}{1 + \frac{3 \rho^2}{2}} \Bigg).
\end{equation}
The absolute value of the 3-momentum
\begin{equation}
P^\ast(Q^2)=\frac{\sqrt{(M_\Delta^2 \, - \, M_N^2 \, - \, Q^2)^2 \, + \,
4 \, M_\Delta^2 \, Q^2}}{2 M_\Delta}
\end{equation}
of either the nucleon or the virtual photon is evaluated in the
$\Delta$-rest frame.

(b) The three-quark counterterm (CT) [Fig.1(b)] results in the expression:
\eq\label{A12CT_0}
A_{1/2}(Q^2)\bigg|_{CT} = (\hat Z^0 - 1) A_{1/2}(Q^2)\bigg|_{3q}^{LO}.
\en

(c) The meson-cloud diagram (MC) [Fig.1(c)] yields:
\eq
A_{1/2}(Q^2)\bigg|_{MC} &=& -\frac{3}{200}
\frac{e\; P^\ast(Q^2)}{\sqrt{2\omega_{\gamma} }}
\Bigg( \frac{g_A}{\pi F} \Bigg)^2
\int\limits^\infty_0 dp p^4 \int\limits^1_{-1} dx (1 - x^2) \\
&\times& {\cal F}_{\pi N N}(p^2,Q^2,x) t_N(p^2,Q^2,x)\bigg|_{MC},
\nonumber
\en
where
\eq
& &{\cal F}_{\pi N N}(p^2,Q^2,x) =
F_{\pi N N}(p^2) F_{\pi N N}(p^2_+)\, , \\
& &t_N(p^2,Q^2,x)\bigg|_{MC} = 2 D_{\pi}(p^2,Q^2,x) + D_{K}(p^2,Q^2,x)\, ,
\nonumber\\
& &D_{\Phi}(p^2,Q^2,x) = \frac{1}{w^2_{\Phi}(p^2)w^2_{\Phi}(p_+^2)}\, ,
\nonumber\\
& &p_\pm^2 = p^2 + Q^2  \pm 2 p \sqrt{Q^2} x \, .
\nonumber
\en

(d) For the vertex-correction diagram (VC) [Fig.1(d)] we obtain:
\eq
A_{1/2}(Q^2)\bigg|_{VC} &=& -\frac{1}{200}
\frac{e \; P^\ast(Q^2)}{\sqrt{2 \omega_{\gamma}}}
\Bigg( \frac{g_A}{\pi F} \Bigg)^2 {\rm exp}
\Big( -\frac{Q^2R^2}{4} \Big) \frac{\rho R}{1 + \frac{3 \rho^2}{2}}\\
&\times&\int\limits^\infty_0
dp p^4 F^{2}_{\pi N N}(p^2) t_N(p^2)\bigg|_{VC} \nonumber
\en
where
\eq
& &t_N (p^2)\bigg|_{VC} = W_{\pi}(p^2) -\frac{1}{3} W_{\eta}(p^2), \\
& &W_{\Phi}(p^2) = \frac{1}{w^{3}_{\Phi}(p^2)}. \nonumber
\en

(e) And for the meson-in-flight diagram (MF) [Fig.1(e)] we get:
\eq
A_{1/2}(Q^2)\bigg|_{MF} &=& -\frac{9}{200}
\frac{e \; P^\ast(Q^2)}{\sqrt{2 \omega_{\gamma}}}
\Bigg( \frac{g_A}{\pi F} \Bigg)^2 \int\limits^\infty_0 dp
p^4 \int^{1}_{-1} dx (1-x^2) \\
&\times& {\cal F}_{\pi N N}(p^2,Q^2,x) D_{\pi}(p^2,Q^2,x). \nonumber
\en

When including excited states in the quark propagator, the analytical
results for the LO three-quark diagram and the meson-in-flight contribution
obviously still remain the same. In turn the following contributions must be
extended.

(a) In the three-quark NLO expression the renormalized quark mass has to
be replaced accordingly
\begin{equation}\label{A12NLO_F}
A_{1/2}(Q^2)\bigg|^{NLO}_{3q} = A_{1/2}(Q^2)\bigg|^{LO}_{3q}
\hat m^r_F\Bigg(\frac{\rho R}{1 + \frac{3 \rho^2}{2}} \Bigg)
\Bigg(\frac{Q^2 R^2}{4} - \frac{2 -
\frac{3 \rho^2}{2}}{1 + \frac{3 \rho^2}{2}} \Bigg).
\end{equation}

(b) For the three-quark counterterm (CT) the appropriate renormalization constant
has to be inserted
\eq\label{A12CT_F}
A_{1/2}(Q^2)\bigg|_{CT} = (\hat Z^F - 1) A_{1/2}(Q^2)\bigg|_{3q}^{LO} .
\en

(c) For the meson-cloud diagram (MC) we obtain the full expression
\eq
A_{1/2}(Q^2)\bigg|_{MC} &=& -\frac{1}{6}
\frac{e \; P^\ast(Q^2)}{\sqrt{2 \omega_{\gamma}}}
\Bigg( \frac{1}{\pi F} \Bigg)^2 \int\limits^\infty_0 dp p^3
\int\limits^1_{-1} dx \frac{1-x^2}{\sqrt{p^2_+}}
\nonumber\\
&\times& \sum\limits_\alpha
{\cal F}_{\alpha;MC}(p^2,Q^2,x) t(p^2,Q^2,x)|_{\alpha;MC} \nonumber
\en
where the sum runs over the index $\alpha$, labelling the quantum numbers
of the intermediate quark states (ground and excited states).
We use the definitions
\begin{equation}
t(p^2,Q^2,x)|_{\alpha;MC} = 2 {\cal D}^{\pi}_{\alpha}(p^2,Q^2,x) +
{\cal D}^{K}_{\alpha}(p^2,Q^2,x)
\end{equation}
and
\begin{equation}
{\cal D}^{\phi}_{\alpha} = \frac{1 + [\Delta
{\cal E}_{\alpha}/(w_{\phi}(p^2_+) + w_{\phi}(p^2))]}
{w_{\phi}(p^2) w_{\phi}(p^2_+)(w_{\phi}(p^2) + \Delta {\cal E}_{\alpha})
(w_{\phi}(p^2_+) + \Delta {\cal E}_{\alpha})},
\end{equation}
where in addition we introduce the function
${\cal F}_{\alpha;MC}(p^2,Q^2,x)=F_{\alpha}(p^2_+)
F^{\dagger}_{\alpha}(p^2)$ with the vertex form factor $F_{\alpha}(p^2)$
of Eq. (\ref{F-ex}).

(d) For the vertex-correction diagram (VC) inclusion of excited states results in
\begin{equation}
A_{1/2}(Q^2)\bigg|_{VC} = \sum\limits_{\beta,\alpha}\frac{I_{\beta
\alpha}(Q^2)}{18} \frac{e \; P^\ast(Q^2)}{\sqrt{2 \omega_{\gamma}}}
\Bigg(\frac{1}{\pi F}\Bigg)^2 \int\limits^\infty_0 dp p^2 {\cal F}_{\beta
\alpha;VC}(p^2) t(p^2)|_{\beta \alpha;VC},
\end{equation}
where
\begin{equation}
t(p^2)|_{\beta \alpha;VC} = {\cal W}^{\pi}_{\beta \alpha}(p^2)
-\frac{1}{3} {\cal W}^{\eta}_{\beta \alpha}(p^2),
\end{equation}
\begin{equation}
{\cal W}^{\phi}_{\beta \alpha}(p^2) =
\frac{1}{w_{\phi}(p^2)(w_{\phi}(p^2) + \Delta{\cal E}_{\beta})
(w_{\phi}(p^2) + \Delta{\cal E}_{\alpha})}
\end{equation}
and ${\cal F}_{\beta \alpha;VC}=F_{\beta}(p^2)
F^{\dagger}_{\alpha}(p^2)$. We also define
\eq
I_{\beta \alpha}(Q^2) &=& 2 N_{\beta} N_{\alpha}
\frac{\partial}{\partial Q^2}\int\limits^\infty_0 dr \,r\,
(g_{\beta}(r) f_{\alpha}(r) + g_{\alpha}(r) f_{\beta}(r))
\nonumber \\
&\times& \int\limits_{\Omega} \, d\cos\theta \, d\phi \,
e^{i \sqrt{Q^2} r {\rm cos} \theta}
C_{\beta} Y_{l_{\beta}\,0}(\theta,\phi)
C_{\alpha} Y_{l_{\alpha}\,0}(\theta,\phi)\, ,
\en
where $l_{\beta}$ and $l_{\alpha}$ are the orbital quantum numbers of
the intermediate states $\beta$ and $\alpha$, respectively.

The result for the $Q^2$-dependence of the helicity amplitude
$A_{1/2}(Q^2)$, when truncating the quark propagator to the ground state,
is indicated in Fig. 2. Thereby, we also list the individual contributions
of the different diagrams of Fig. 1, which add up coherently.
The leading order three quark diagram dominates the prediction for $A_{1/2}$,
whereas meson cloud corrections add about $30\%$ to the total result.
Here, both the meson-in-flight and the meson-cloud diagrams give the
largest contribution.
In a next step we include the intermediate excited quark
states with quantum numbers $1p_{1/2}$, $1p_{3/2}$, $1d_{3/2}$,
$1d_{5/2}$, and $2s_{1/2}$, that is excitations up to 2 $\hbar \omega$,
in the propagator. The resulting effect on $A_{1/2}$ is given in
Fig. 3. We explicitly indicate the additional terms, which are
solely due to the excited states, whereas the ground state propagator
result is contained in the curve denoted by TOTAL(GS).
Although higher lying state contributions are suppressed relative to
the ground state one, they still have a noticeable effect on $A_{1/2}$
for $Q^2 < 0.5$ ${\rm GeV^2}$ at the order of $15\%$.
For completeness we give the full result for both transverse
helicity amplitudes $A_{1/2}(Q^2)$ and $A_{3/2}(Q^2)$ in Fig. 4.
The amplitudes fulfil the relation
$A_{3/2}(Q^2) = \sqrt{3} A_{1/2}(Q^2)$ for all $Q^2$ values.
\par
Recently, in the
framework of large-$N_c$ QCD \cite{Jenkins} it was shown that
the ratio $A_{3/2}/A_{1/2}$ is mostly saturated by the
naive $SU_6$ quark model result $A_{3/2}/A_{1/2} = \sqrt{3}$
\cite{Close}. Deviations from this standard result are due to higher
order corrections with $A_{3/2}/A_{1/2} = \sqrt{3} + O(1/N_c^2)$
\cite{Jenkins}. We evaluate the helicity amplitudes at one-loop
or equivalently to the order of accuracy $o(1/F^2,\hat m,m_s)$, where
$F \sim \sqrt{N_c}$. Therefore, to get a nontrivial deviation from
the $SU_6$ result the formalism has to be extended up to two loops or up
to order $O(1/F^4) \sim O(1/N_c^2)$. The standard relation
between the helicity amplitudes obtained here, which
is consistent with large-$N_c$ considerations, is for example not
present in the cloudy bag model \cite{KE,Bermuth,Lu}. Latter model,
which is conceptually close to our approach, uses in the calculation
of one-loop diagrams non-degenerate nucleon and delta masses. This
would correspond in our counting scheme to the order of accuracy
$O(1/F^4)$, although the complete diagrams to this order are not evaluated.
Based on the arguments of the large-$N_c$ analysis it is not surprising
that the cloudy bag is able
to generate a deviation from the $\sqrt{3}$ result and hence produces
a non-vanishing $E2/M1$ ratio.

For comparison with data we turn to the results for the helicity
amplitudes at the real-photon point with $Q^2 =0$.
In Table I we list the numerical values for the complete set
of Feynman diagrams, again indicating separately the contributions
of ground and excited states in the quark propagator.
Comparison of our results to other model calculations are presented
in Table III. Our final results for $A_{1/2}(Q^2 =0)$ and $A_{3/2}
(Q^2 =0)$ are in rather decent agreement with the data. As evident
from Table I, meson cloud corrections play a decisive role in
explaining the large deviation from the result of the impulse,
that is three-quark core, approximation.
Pion contributions play
the dominant role in the meson corrections as evident
from Table III, where we list the individual contributions of the
octet mesons to the sizable terms generated by the meson-cloud and
vertex-correction diagrams. The suppression of $K$ and $\eta$ loops
is due to the large corresponding meson masses occurring in the
denominators.
The relative contribution of $K$ and $\eta$ mesons with respect to $\pi$
is between $8-10\%$ in the amplitude, as can be naively expected
from the ratio of meson masses $(m_{\pi}/m_{K,\eta})^2$ which also
is roughly $8\%$.
Also, the improved
treatment of the quark propagator by including higher excitations
tends at least phenomenologically to be required by the data.
Again, to obtain a non-vanishing value for the $E2/M1$ ratio,
higher order or two-loop corrections have to be considered.

To complete our set of predictions we also indicate the results
for the radiative transition $\Delta^+ \to p \gamma$.
For the decay width as based on Eq. (\ref{dwidth}) we obtain
$ \Gamma(\Delta^+ \to p \gamma) = 0.55 \pm 0.03$ MeV.
Using the experimental value $\Gamma_{total}(\Delta^+)=111.2$ MeV for the
total decay width, we deduce the branching ratio
\eq
BR(\Delta^+ \to p \gamma) = (0.47-0.52)\%
\en
and in similar fashion for the partial decay branching ratios
for helicity 1/2 and 3/2
\eq
BR_{1/2} = (0.12-0.13)\% \, \, {\rm and}\, \,  BR_{3/2} = (0.35-0.39)\%.
\en
These results are again in good agreement with the experimental data of
\cite{PDG}: ${\rm BR}(\Delta^+ \to p \gamma) = (0.52 - 0.60) \%$,
${\rm BR}_{1/2} = (0.11 - 0.13) \%$ and
${\rm BR}_{3/2} = (0.41 - 0.47) \%$.

\section{Summary}

In summary, we have calculated the transverse helicity amplitudes
for the transition $\gamma N \to \Delta$ in the perturbative chiral
quark model. Meson cloud corrections are crucial to explain
the magnitude of the helicity amplitudes at the real-photon point. 
These meson cloud effects were found to be important not
only in $\gamma N \to \Delta$ transition but also in weak pion production
reactions~\cite{Sato}, in vector, axial-vector
and strong $NN$ and $N\Delta$ couplings~\cite{Hemmert} and in data on
light meson photoproduction~\cite{Kamalov}.
     
We demonstrated furthermore that in the context of the PCQM excited
quark states in loop diagrams play an important role at the level
of $15\%$ to fully account for the measurements.
Because at one-loop we work at the order of accuracy
$o(1/F^2,\hat m,m_s)$ or equivalently at $o(1/N_c)$, a deviation
from the standard ratio of $A_{3/2}/A_{1/2} = \sqrt{3}$, consistent
with large-$N_c$ arguments~\cite{Jenkins} cannot be obtained.
Hence, we also predict a vanishing value for the $E2/M1$ ratio.

A next step will be to explore the effect of two-loop
diagrams on the helicity amplitudes, which could possibly explain
the non-vanishing $E2/M1$ ratio.
In view of the important role of intermediate excited quark
states in loop diagrams,
it would be also interesting to check their influence on other baryon
observables.

\vspace*{.5cm}

{\bf Acknowledgements}

\vspace*{.1cm}

\noindent
K.P. thanks the Development and Promotion of Science and Technology
Talent Project (DPST), Thailand for financial support. S.C.
acknowledge the support of Thailand Research Fund (TRF,Grant No. RGJ PHD/00165/2541).
This work was supported by the Deutsche Forschungsgemeinschaft (DFG)
under contracts FA67/25-1 and GRK683.

\appendix
\section{Solutions of the Dirac equation for the effective potential}

In this section we indicate the solutions to the Dirac equation
with the effective potential $V_{\rm eff}(r) = S(r) + \gamma^0
V(r)$. The scalar $S(r)$ and time-like vector $V(r)$ parts are
given by
\eq\label{poten}
S(r)&=& M_1 + c_1 r^2,
\nonumber \\
V(r) &=& M_2 + c_2 r^2,
\en
with the particular choice
\begin{equation}
M_1 = \frac{1 \, - \, 3\rho^2}{2 \, \rho R} , \hspace*{1cm}
M_2 = {\cal E}_0 - \frac{1 \, + \, 3\rho^2}{2 \, \rho R} , \hspace*{1cm}
c_1 \equiv c_2 =  \frac{\rho}{2R^3} .
\end{equation}
The quark wave function $u_{\alpha}(\vec r)$ in state $\alpha$
with eigenenergy ${\cal E}_{\alpha}$ satisfies the Dirac equation
\begin{equation}\label{dirac}
[-i \vec \alpha \vec \nabla +\beta S(r) + V(r) - {\cal E}_{\alpha}]
u_{\alpha} (\vec r) = 0.
\end{equation}
Solutions of the Dirac spinor $u_{\alpha}(\vec r)$ to Eq. (\ref{dirac}) can
be written in the form \cite{Tegen}
\begin{equation}
u_{\alpha}(\vec r)= N_{\alpha}
\left( \begin{array}{c} g_{\alpha}(r) \\ i \vec \sigma
\cdot \hat r f_{\alpha}(r) \end{array} \right) {\cal Y}_{\alpha}(\hat r)
\chi_f \chi_c .
\end{equation}
For the particular choice of potential the radial functions g and f
satisfy the form
\begin{equation}
g_{\alpha}(r) = \bigg( \frac{r}{R_{\alpha}} \bigg)^l
L^{l+1/2}_{n-1}\bigg( \frac{r^2}{R^2_{\alpha}} \bigg) e^{-\frac{r^2}{2
R^2_{\alpha}}},
\end{equation}
where for $j=l+\frac{1}{2}$
\begin{equation}
f_{\alpha}(r)=\rho_{\alpha} \bigg(\frac{r}{R_{\alpha}}\bigg)^{l+1}
\bigg[L^{l+3/2}_{n-1}(\frac{r^2}{R^2_{\alpha}}) +
L^{l+3/2}_{n-2}(\frac{r^2}{R^2_{\alpha}})\bigg] e^{-\frac{r^2}{2 R^2_{\alpha}}},
\end{equation}
and for $j=l-\frac{1}{2}$
\begin{equation}
f_{\alpha}(r)=-\rho_{\alpha} \bigg(\frac{r}{R_{\alpha}}\bigg)^{l-1}
\bigg[(n+l-\frac{1}{2})L^{l-1/2}_{n-1}(\frac{r^2}{R^2_{\alpha}})
+ nL^{l-1/2}_{n}(\frac{r^2}{R^2_{\alpha}})\bigg] e^{-\frac{r^2}
{2 R^2_{\alpha}}}.
\end{equation}
The label $\alpha=(nljm)$ characterizes the state with
principle quantum number $n=1,2,3,...$, orbital angular momentum $l$,
total angular momentum $j=l\pm \frac{1}{2}$ and projection $m$.
Due to the quadratic nature of the potential the radial
wave functions contain the associated
Laguerre polynomials $L^{k}_{n}(x)$ with
\begin{equation}
L^{k}_{n}(x)=\sum^{n}_{m=0} (-1)^m \frac{(n+k)!}{(n-m)!(k+m)!m!} x^m.
\end{equation}
The angular dependence (${\cal Y}_{\alpha}(\hat r) \equiv
{\cal Y}_{lmj}(\hat r)$) is defined by
\begin{equation}
{\cal Y}_{lmj}(\hat r)=\sum_{m_l,m_s} (l m_l \frac{1}{2} m_s | j m)
Y_{l m_l}(\hat r) \chi_{\frac{1}{2} m_s}
\end{equation}
where $Y_{l m_l}(\hat r)$ is the usual spherical harmonic.
Flavor and color part of the Dirac spinor are represented
by $\chi_f$ and $\chi_c$, respectively.

The normalization constant is obtained from the condition
\begin{equation}
\int\limits^\infty_0 d^3 \vec r u^{\dagger}_{\alpha}(\vec r)
u_{\alpha}(\vec r) = 1
\end{equation}
which results in
\begin{equation}
N_{\alpha}=\bigg[ 2^{-2(n+l+1/2)} \pi^{1/2} R^3_{\alpha}
\frac{(2n+2l)!}{(n+l)!(n-1)!}\{1 + \rho^2_{\alpha}
(2n + l -\frac{1}{2})\} \bigg]^{-1/2},
\end{equation}
The two coefficients $R_{\alpha}$ and $\rho_{\alpha}$ are of
the form
\eq
R_{\alpha} &=& R(1 + \Delta {\cal E}_{\alpha} \rho R)^{-1/4},\\
\rho_{\alpha} &=& \rho \bigg( \frac{R_{\alpha}}{R} \bigg)^3
\en
and are related to the Gaussian parameters $\rho$, $R$ of
Eq. (\ref{Gaussian_Ansatz}).
The quantity $\Delta {\cal E}_{\alpha} ={\cal E}_{\alpha}- {\cal E}_0$
is the difference between the energy
of state $\alpha$ and the ground state.
$\Delta {\cal E}_{\alpha}$ depends on the quantum numbers $n$ and $l$
and is related to the parameters $\rho$ and $R$ by
\begin{equation}
(\Delta {\cal E}_{\alpha} + \frac{3 \rho}{R})^2
(\Delta {\cal E}_{\alpha} + \frac{1}{\rho R}) =
\frac{\rho}{R^3} (4 n +2 l -1)^2.
\end{equation}

\newpage
\centerline{TABLES}

\vspace*{1cm}

{\bf Table I.} Contributions of the individual diagrams
to the transverse helicity amplitudes for
$Q^2 = 0$ (in units of $10^{-3}\,{\rm GeV}^{-1/2}$).
Results for inclusion of ground (GS) and excited states (ES) in
the quark propagator are indicated separately.

\vspace*{.2cm}

\begin{center}
\begin{tabular}{lcc}
\hline \hline
  & $A_{1/2}(Q^2=0)$ & $A_{3/2}(Q^2=0)$ \\ \hline \hline
{\bf GS quark propagator} & & \\
3q-core & & \\
-LO     & -69.7 $\pm$ 5.9& -120.7 $\pm$ 10.2\\
-NLO    & -8.6 $\pm$ 1.2& -14.9 $\pm$ 2.1\\
Counter-term & 8.2 $\pm$ 1.1& 14.2 $\pm$ 1.9\\
Meson-cloud & -16.7 $\pm$ 2.6& -28.9 $\pm$ 4.5\\
Vertex-correction & -0.7 $\pm$ 0.1& -1.2 $\pm$ 0.1\\
Meson-in-flight & -23.0 $\pm$ 3.4& -39.8 $\pm$ 5.9\\
\hline
Total(GS) & -110.5 $\pm$ 0.3 & -191.3 $\pm$ 0.5 \\
\hline
{\bf ES quark propagator} & & \\
NLO & -10.3 $\pm$ 1.1& -17.8 $\pm$ 1.9\\
Counter-term & 4.9 $\pm$ 0.6& 8.5 $\pm$ 1.0\\
Meson-cloud & -13.5 $\pm$ 2.5& -23.4 $\pm$ 4.3\\
Vertex-correction &-0.7 $\pm$ 0.1& -1.2 $\pm$ 0.1\\
\hline
Total(ES) & -19.6 $\pm$ 3.1 & -33.9 $\pm$ 5.3 \\
\hline \hline
Total=Total(GS)+Total(ES) & -130.1 $\pm$ 3.4& -225.2 $\pm$ 5.8\\
Experiment \cite{PDG} & -135 $\pm$ 6 & -255 $\pm$ 8 \\
\hline \hline
\end{tabular}
\end{center}

\newpage

{\bf Table II.} Total result for the helicity amplitudes $A_{1/2}(Q^2 =0)$
and $A_{3/2}(Q^2 =0)$ in comparison to other theoretical models
(in units of $10^{-3}\;{\rm GeV}^{-1/2}$). The cloudy bag model results are
taken from the values at a typical bag radius $R=0.8$ fm.

\vspace*{.2cm}

\begin{center}
\begin{tabular}{lcc}
\hline \hline
  & $A_{1/2}(Q^2=0)$ & $A_{3/2}(Q^2=0)$ \\ \hline \hline
NRQM \cite{Buchmann_EM} & -90.9 & -181.9 \\
Cloudy bag model \cite{Lu} & -128 & -222 \\
Relativistic quark potential model \cite{Dong} & -147 & -277 \\
PCQM   & -130.1 $\pm$ 3.4 & -225.2 $\pm$ 5.8 \\
\hline
Experiment \cite{PDG} & -135 $\pm$ 6 & -255 $\pm$ 8 \\
\hline \hline
\end{tabular}
\end{center}

\newpage

{\bf Table III.} Absolute contributions of $\pi$, $K$, and $\eta$ to
$A_{1/2}(Q^2=0)$ for the meson-cloud (MC) and vertex-correction (VC)
diagrams in units of $10^{-3}\;{\rm GeV}^{-1/2}$.

\vspace*{.2cm}
\begin{center}

\begin{tabular}{lcccc}
\hline \hline
 & $A_{1/2}({\pi})$ & $A_{1/2}({K})$ & $A_{1/2}({\eta})$ & Total \\
\hline
{\bf GS quark propagators} & & & & \\
MC & -15.3 & -1.4 & - & -16.7 \\
VC & -0.73 &  - & 0.06 & -0.67 \\
{\bf ES quark propagators} & & & & \\
MC & -12.4 & -1.1 & - & -13.5 \\
VC & -0.80 & - & 0.08 & -0.72\\
\hline \hline
\end{tabular}
\end{center}

\newpage

%%%%%%%%%%%%%%%%%%%%%%%%%%%%%%%%%%%%%%%%%%%%%%%%%
%            FIGURES
%%%%%%%%%%%%%%%%%%%%%%%%%%%%%%%%%%%%%%%%%%%%%%%%%%

\begin{figure}[t]
\noindent Fig.1: Diagrams contributing to the transverse helicity amplitudes:

\noindent 3q-core (1a), counter term (CT) (1b), meson-cloud (MC) 
(1c), vertex-correction (VC) (1d), 
and meson-in-flight (MF) diagram (1e).

\vspace*{1cm}
\noindent Fig.2: $Q^2$-dependence of the transverse helicity amplitude 
$A_{1/2}(Q^2)$ for the case where the quark propagator is truncated to
the ground state (GS) contribution. Legend: [3q-LO] -  3q-diagram
(leading order); [3q-NLO(GS)] - 3q-diagram 
(next-to-leading-order); [CT(GS)] - counter term;
[MC(GS)] - meson-cloud diagram; 
[VC(GS)] - vertex-correction diagram; [MF] -
meson-in-flight diagram; Total(GS) - total result.

\vspace*{1cm}
\noindent Fig.3: Same as Fig. 2 but now for the case where excited 
quark states (ES) are included in the loop diagrams.
Legend: [Total(GS)] - total result for the ground state quark propagator;
[3q-NLO(ES)] - excited states in the 3q-NLO diagram ; [CT(ES)] -
counter term; [MC(ES)] - meson cloud diagram; 
[VC(ES)] - vertex correction diagram; Total = Total(GS)+3q-NLO(ES)+CT(ES)+MC(ES)+VC(ES) - total result.

\vspace*{1cm}
\noindent Fig.4: Total result for the transverse helicity amplitudes
$A_{1/2}(Q^2)$ and $A_{3/2}(Q^2)=\sqrt{3} A_{1/2}(Q^2)$.

\end{figure}

\newpage

\begin{figure}
\centering{\
\epsfig{figure=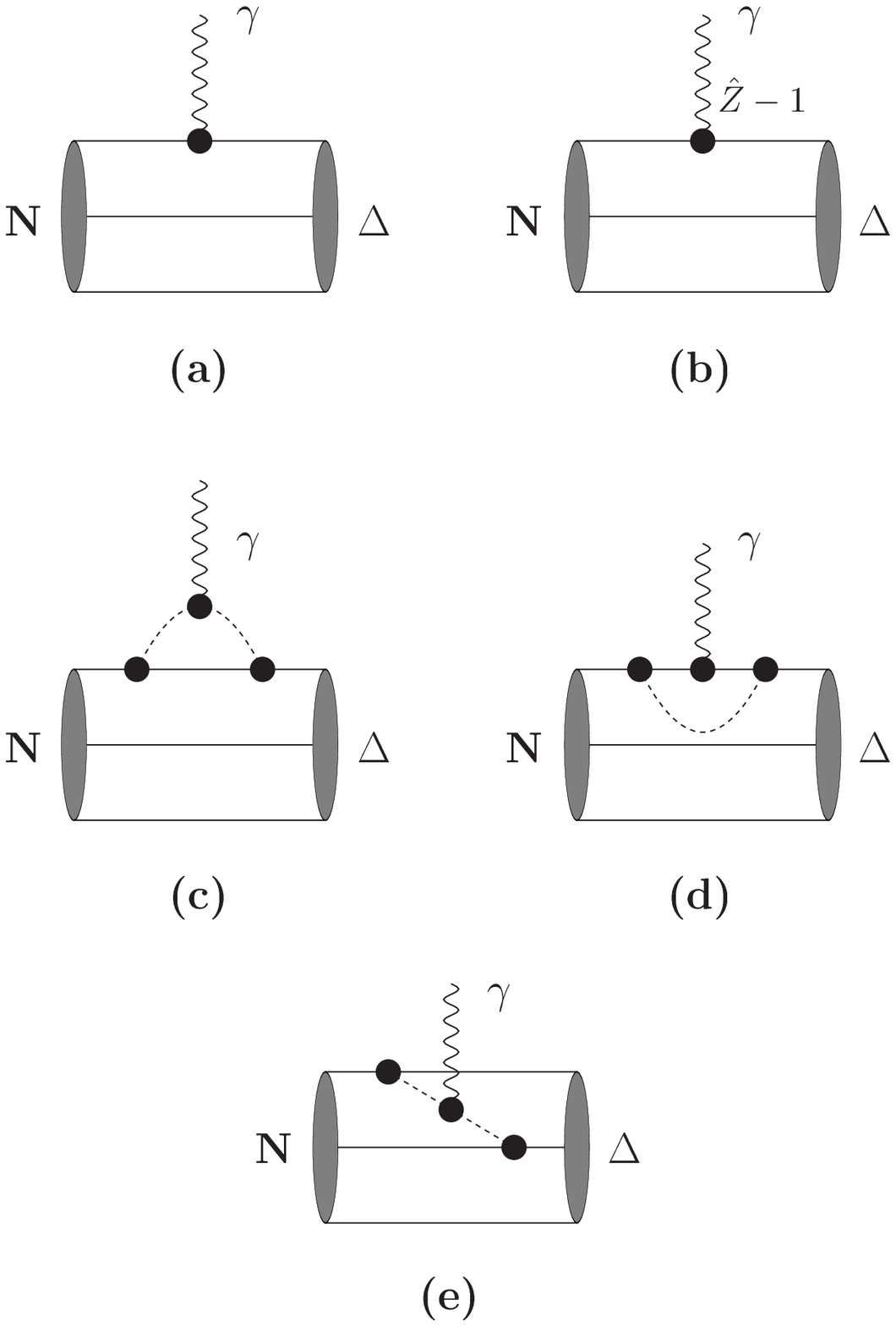,height=21cm}}
\end{figure}

\vspace*{-2cm}

\centerline{\bf Fig.1}

\newpage

\begin{figure}
\centering{\
\epsfig{figure=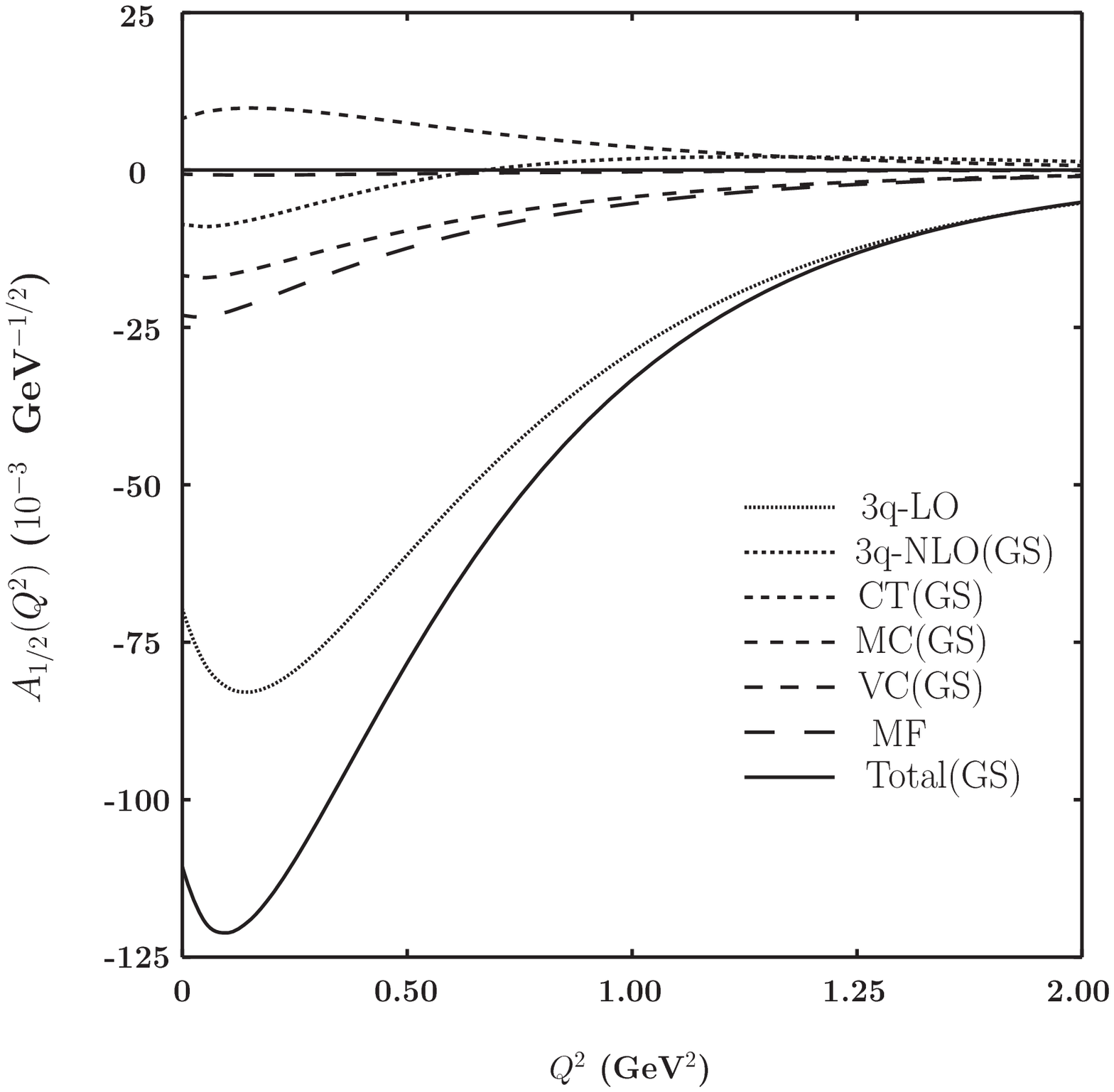,height=21cm}}
\end{figure}

\vspace*{-1cm}

\centerline{\bf Fig.2}

\newpage

\begin{figure}
\centering{\
\epsfig{figure=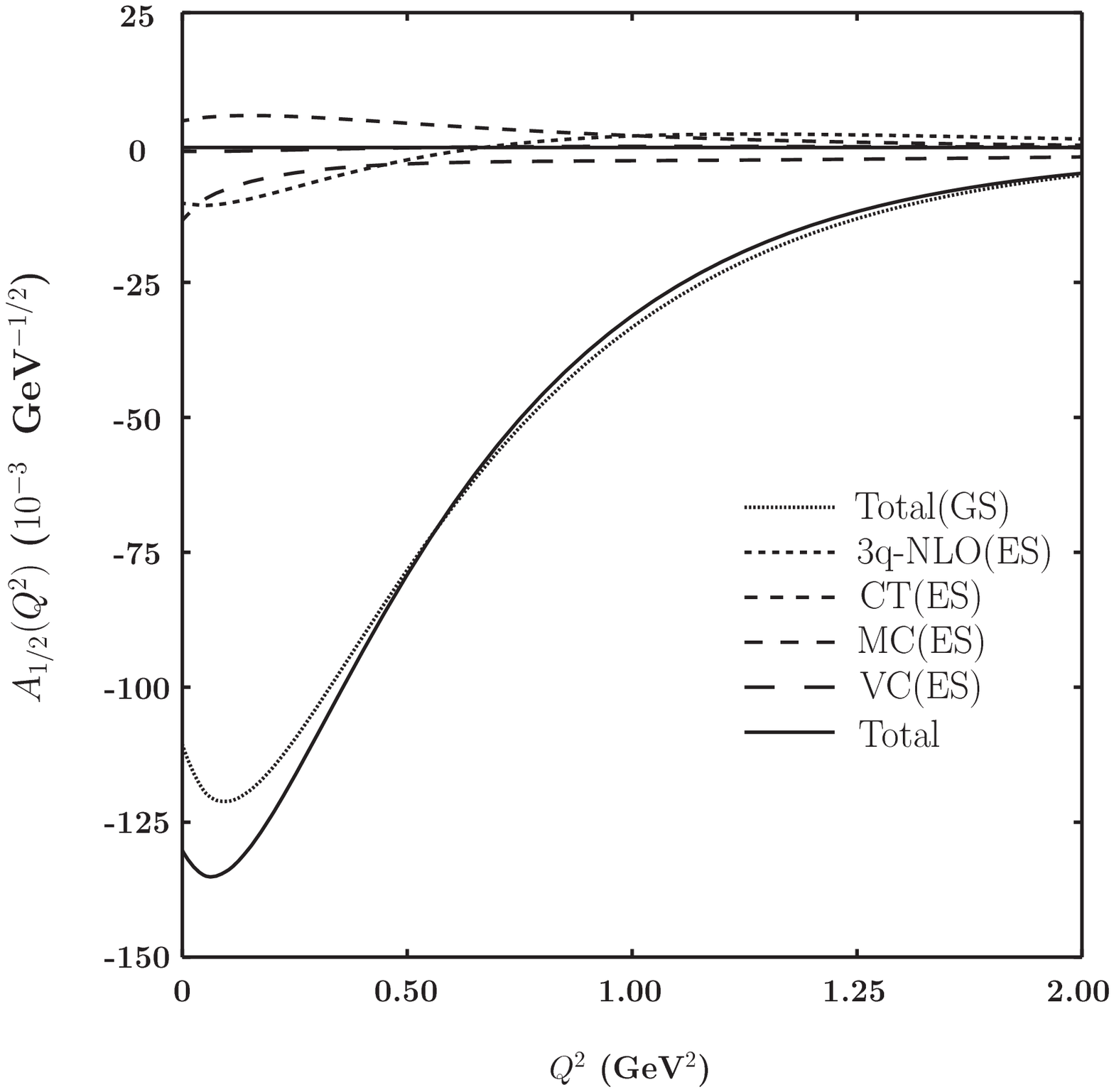,height=21cm}}
\end{figure}

\vspace*{-1cm}

\centerline{\bf Fig.3}

\newpage

\begin{figure}
\centering{\
\epsfig{figure=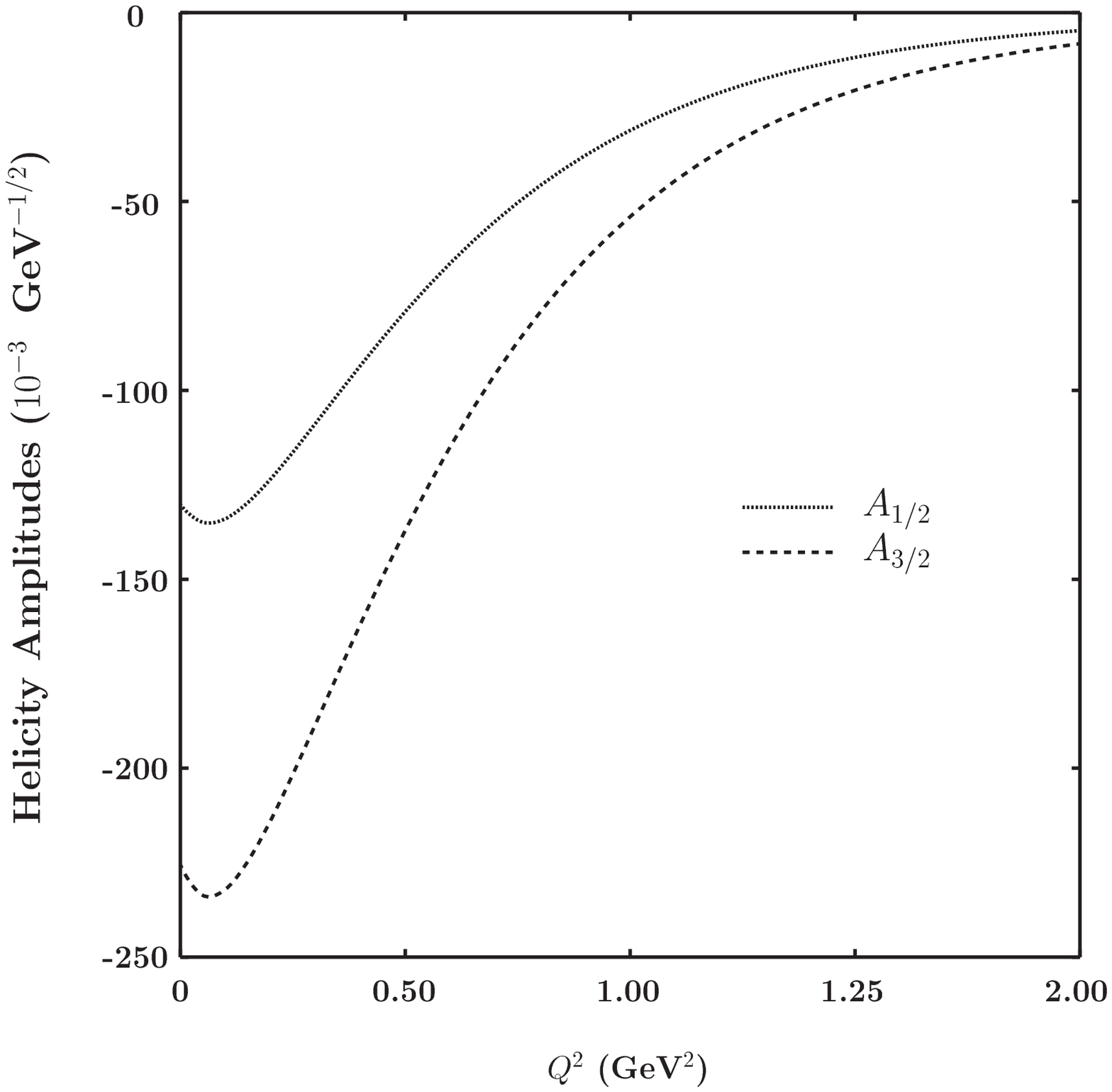,height=21cm}}
\end{figure}

\vspace*{-1cm}

\centerline{\bf Fig.4}


\begin{thebibliography}{99}
\bibitem{Copley}L.~A.~Copley, G.~Karl, and E.~Obryk,
Nucl. Phys. {\bf B13}, 303 (1969).
\bibitem{PDG}K.~Hagiwara {\it et al.},
Phys.\ Rev.\ D {\bf 66}, 010001 (2002).
\bibitem{Jenkins}E.~Jenkins, X.~Ji, A.~V.~Manohar,
Phys.\ Rev.\ Lett. {\bf 89}, 242001 (2002).
\bibitem{IsgurKarl}
N.~Isgur, G.~Karl and R.~Koniuk,
Phys.\ Rev.\ D {\bf 25}, 2394 (1982).
\bibitem{Gershtein}
S.~S.~Gershtein and G.~V.~Jikia,
Sov.\ J.\ Nucl.\ Phys.\  {\bf 34}, 870 (1981)
[Yad.\ Fiz.\  {\bf 34}, 1566 (1981)].
\bibitem{Drechsel}
D.~Drechsel and M.~M.~Giannini, 
Phys.\ Lett.\ B {\bf 143}, 329 (1984).
\bibitem{Buchmann_EM}A.~J.~Buchmann, E.~Hern\'{a}ndez,
and A.~Faessler, Phys. Rev. C {\bf 55}, 448 (1997).
\bibitem{Buchmann_EC}A.~J.~Buchmann, Z. Naturforsch. {\bf 52a},
877 (1997).
\bibitem{Isgur}N.~Isgur and R.~Koniuk,
Phys. Rev. D {\bf 21}, 1868 (1980).
\bibitem{Giannini} M.~M.~Giannini,
Rep. Prog. Phys. {\bf 54}, 453 (1990).
\bibitem{KE}G.~Kalbermann and J.~M.~Eisenberg,
Phys. Rev. D {\bf 28}, 71 (1983).
\bibitem{Bermuth}K.~Bermuth, D.~Dreschsel, L.~Tiator, and
J.~B.~Seaborn, Phys. Rev. D {\bf 37}, 89 (1986).
\bibitem{Lu}D.~H.~Lu, A.~W.~Thomas, and A.~G.~Williams,
Phys. Rev. C {\bf 55}, 3108 (1997).
\bibitem{Dong}Y.~B.~Dong, K.~Shimizu, A.~Faessler,
Nucl. Phys. {\bf A689}, 889 (2001).
\bibitem{Sato}T.~Sato and T.~-S.~H.~Lee, Phys. Rev. C {\bf 63}, 055201 (2001);
T.~Sato, D.~Uno and T.~-S.~H.~Lee, nucl-th/0303050.
\bibitem{Hemmert}T.~R.~Hemmert and B.~R.~Holstein,
Phys. Rev. D {\bf 51}, 158 (1995).
\bibitem{Kamalov}S.~A.~Kamalov and S.~N.~Yang, Phys. Rev. Lett.
{\bf 83}, 4494 (1999);
S.~A.~Kamalov, S.~N.~Yang, D.~Drechsel, O.~Hanstein,
and L.~Tiator,  Phys. Rev. C {\bf 64}, 032201(R) (2001).
\bibitem{Theberge}S.~Th\'{e}berge, A.~W.~Thomas, and G.~A.~Miller,
Phys. Rev. D {\bf 22}, 2838 (1980); Phys. Rev. D {\bf 24} (1981) 216.
\bibitem{Oset} E. Oset, R. Tegen and W. Weise, Nucl. Phys.
A {\bf 426} (1984) 456; R. Tegen, Ann. Phys. {\bf 197} (1990) 439.
\bibitem{Chin} S. A. Chin, Nucl. Phys. A {\bf 382} (1982) 355.
\bibitem{Gutsche}T.~Gutsche and D.~Robson, Phys. Lett. B {\bf 229}
(1989) 333; T.~Gutsche, Ph. D. Thesis, Florida State University,
1987 (unpublished).
\bibitem{PCQM1} V. E. Lyubovitskij, T. Gutsche and A. Faessler,
Phys. Rev. C {\bf 64}, 065203 (2001).
\bibitem{PCQM2} V. E. Lyubovitskij, T. Gutsche, A. Faessler and
E. G. Drukarev, Phys. Rev. D {\bf 63}, 054026 (2001).
\bibitem{PCQM3} V.E. Lyubovitskij, Th. Gutsche,
A. Faessler, and R. Vinh Mau,  Phys. Lett. B {\bf 520}, 204 (2001);
Phys. Rev. C {\bf 65}, 025202 (2002).
\bibitem{PCQM4} V.E. Lyubovitskij, P. Wang, Th. Gutsche, and
A. Faessler,  Phys. Rev. C {\bf 66}, 055204 (2002).
\bibitem{Gasser_Sainio} J.~Gasser,M.~E.~Sainio, and A.~B.~ $\rm \check{S}$varc,
Nucl. Phys. {\bf B307}, 779 (1988).
\bibitem{Gasser_Leutwyler}J.~Gasser and H.~Leutwyler,
Phys. Rep. {\bf 87}, 77 (1982).
\bibitem{Close}F.~E.~Close, An Introduction to Quarks and Partons
(Academic Press, New York, 1979).
\bibitem{Tegen} R.~Tegen and R.~Brockmann,
Z. Phys. A {\bf 307}, 339 (1982).
\end{thebibliography}
\end{document}